\documentclass[12pt,preprint]{aastex}
\usepackage{graphicx}

\def\mb{\ifmmode {{\rm B_{435}}}\else
                ${\rm B_{435}}$\fi}
\def\mv{\ifmmode {{\rm V_{606}}}\else
                ${\rm V_{606}}$\fi}
\def\mi{\ifmmode {{\rm i_{775}}}\else
                ${\rm i_{775}}$\fi}
\def\mz{\ifmmode {{\rm z_{850}}}\else
                ${\rm z_{850}}$\fi}

\def\mY{\ifmmode {{\rm Y_{105}}}\else
                ${\rm Y_{105}}$\fi}
\def\mJ{\ifmmode {{\rm J_{125}}}\else
                ${\rm J_{125}}$\fi}
\def\mH{\ifmmode {{\rm H_{160}}}\else
                ${\rm H_{160}}$\fi}

\def\PMCMC{$\pi$MC$^{2}$}               
                
\begin{document}
\title{REAL OR INTERLOPER? The Redshift Likelihoods Of $Z>8$\ Galaxies In The Hudf12}
\author{Nor Pirzkal\altaffilmark{1}, Barry Rothberg\altaffilmark{2,3},\\ 
Russell Ryan\altaffilmark{1}, Dan Coe\altaffilmark{1}, 
Sangeeta Malhotra\altaffilmark{4}, James Rhoads\altaffilmark{4}, AND Kai Noeske\altaffilmark{1}}
\altaffiltext{1}{Space Telescope Science Institute, 3700 San Martin Drive, Baltimore, MD21218, USA}
\altaffiltext{2}{Leibniz-Institut f\"{u}r Astrophysik Potsdam (AIP), An der Sternwarte 16, D-14482, Potsdam, Germany}
\altaffiltext{3}{Department of Physics \& Astronomy,George Mason University, MS 3F3, 4400 University Drive, Fairfax, VA 22030, USA}
\altaffiltext{4}{School of Earth And Space Exploration, Arizona State University, Tempe, AZ, 85287-1404, USA}

\begin{abstract}
In the absence of spectra, the technique of fitting model galaxy template spectra to observed photometric
 fluxes has become the workhorse method for determining the redshifts and other properties for high-$z$\ galaxy candidates.  
In 
this paper, we present an analysis of the most recent and possibly most distant galaxies ($z\sim$8 -- 12)
discovered in the Hubble Ultra Deep Field (HUDF) using a more robust method of redshift estimation based on Markov Chain Monte Carlo (MCMC)
fitting, in contrast to the ``best fit'' models obtained using 
simpler $\chi$$^{2}$\ minimization techniques.
The advantage of MCMC fitting is the ability to  accurately estimate the probability 
density function of the redshift, for each object as well as any input model parameters. This makes it possible 
to derive accurate credible intervals by properly marginalizing over all other input model 
parameters.  We apply our method to 13 recently identified sources in the HUDF and show that, despite 
claims based on $\chi$$^{2}$ minimization, none of these sources can be securely ruled 
out as low redshift interlopers ($z<$4) due to the low signal-to-noise of currently available 
observations. There is an average probability of $21\%$\ that these 
sources are low redshift interlopers. 
\end{abstract}


\section{INTRODUCTION}
\indent Over the last 15 yr the ever increasing pace at which new, more sensitive 
detectors and larger telescope apertures have come online has spurred a fast and furious 
race to detect the most distant galaxies in the universe.  Emission features from HII 
regions (such as Ly$\alpha$), have been a useful tool for detecting star-forming 
galaxies at 3 $<z<$ 6 
\citep[corresponding to rest-frame optical wavelengths;][]{Malhotra2001}.  However, at higher redshifts, corresponding to the first several 
hundred million years since the big bang, more and more matter was locked up in neutral 
hydrogen.  At these early epochs the neutral intergalactic medium attenuates 
significant amounts of light from galaxies, including Ly$\alpha$, making detection of 
these systems more difficult.  If Ly$\alpha$ is significantly diminished then one must 
rely on the detection of a continuum break to select candidate galaxies.  Detecting 
galaxies at these redshifts places constraints on the epoch of re-ionization, thought to 
be at z$>$6 \citep{Fan2006,Komatsu2011}.  Detecting and measuring the properties of these 
galaxies is critical to understand what caused the epoch of re-ionization.\\ 
\indent Detecting and measuring the properties of high-$z$\ galaxies is not easy. At 
rest-frame $\lambda$ $<$1216\AA\ the flux from high redshift galaxies can be completely vanquished 
due to the Ly$\alpha$\ forest absorption and when observed in a filter corresponding 
to this wavelength the galaxy ``drops out.''  At $\lambda$ $>$ 1216\AA\ the flux is not 
attenuated and can be detected.  Such breaks become more pronounced with redshift, for 
example at $z>$ 5.7 a break corresponds to $\sim$ 3.4 mag in color 
\citep[e.g.][]{Songaila2002,Hu2004}. Detection at $z>$ 8 is further complicated by the 
fact that the Lyman regime is redshifted into near-infrared wavelengths 
($\lambda$ $>$ 1 $\micron$) where bright atmospheric telluric skylines and strong 
telluric absorption features force ground-based observations to use specific and
restrictive atmospheric windows. {\em Large Hubble Space Telescope (HST)} based spectroscopic surveys such as 
GRAPES and PEARS \citep{Pirzkal2004} have demonstrated the power of low resolution 
spectroscopy to identify high redshift sources \citep{Malhotra2005, Rhoads2009}. However,
until similar large-scale programs are implemented in the near-IR using the WFC3/{\em HST} or {\em James Webb Space Telescope (JWST)}, the
next best method for detecting the most distant galaxies relies on using near-IR photometry obtained
with {\em HST}.
\\
\indent Recently, the race to find the earliest galaxies has relied more and more on the 
use of the near-infrared Wide Field Camera 3 (WFC3) detector 
(1 $<$ $\lambda$ $<$ 1.6 $\micron$) on the  {\it HST}, and in 
particular, broad-band imaging.  Detecting possible 1216\AA\ breaks with 
{\it HST} photometry should be straightforward, assuming one collects a sufficient amount 
of photons.  However, observations must first contend with the possibility of foreground
interlopers. At  $z\sim$ 6 foreground interlopers can be as frequent as 45$\%$ for 
observations with signal-to-noise (S/N) $<$ 5 \citep[e.g.][]{Dickinson2004, Stanway2008}.
Without the direct detection of spectroscopic lines, redshift confirmations must rely
on Spectral Energy Distribution (SED) fitting. \\
\indent One of the first tools used to derive photometric redshifts was the SED fitting 
code {\tt hyperz} which simply maximizes the likelihood by brute-force using a grid of SED templates \citep{Bolzonella2000}.
As noted in \cite{Bolzonella2000}:  ``Both the {\it z}$_{\rm phot}$ and the SED are 
obtained through {\tt hyperz}, together with the best fit parameters ($A_{\rm V}$, 
spectral type, metallicity and age). Because of the degeneracy between these parameters, 
the relevant information shall be the redshift and the rough SED type, in the sense that 
a given object has a ``blue'' or ``red'' continuum at a given $z$, but no reliable 
information can be obtained about the other parameters from broad-band photometry alone.''
Despite this, {\tt hyperz} and other similar $\chi$$^{2}$ minimization SED fitting 
codes \citep[e.g.][]{Thomspon2001,Papovich2001,Labbe2003,Schaerer2005,Mobasher2005,Legos2007,McLure2009} 
have been used to extract more than just redshifts, including these degenerate properties.
All of which is based on simply selecting the template with the smallest $\chi$$^{2}$.\\
\indent This has led to a number of situations in which photometric redshifts and galaxy
properties determined from $\chi$$^{2}$ minimization SED fitting have produced questionable 
claims of high-$z$\ galaxies.  \cite{Pello2004} claimed the discovery of a $z=10$ galaxy
from a combination of deep {\it J, H, K} photometry obtained from the {\it Very Large Telescope (VLT)}, {\it HST}
optical imaging, and {\it J}-band spectroscopy from ISAAC/VLT.  While \cite{Pello2004} claimed
a detection of Ly$\alpha$ from low S/N ISAAC/VLT spectra, the basis for the claim relied
primarily on SED fitting using a modified form of {\tt hyperz} with stacked {\it J, H, K} imaging.  
Followup deep {\it V}-band ({\it V}$_{\rm AB}$ $=$ 27.8 mag) imaging from the VLT 
\citep{Lehnert2005} and deeper {\it H}-band imaging from Gemini/NIRI \citep{Bremer2004} 
failed to detect fluxes predicted from the SED fitting. These imaging results and a re-analysis of
the spectra by \cite{Weatherley2004} led to the conclusion that the object was a spurious 
detection, even though the fits from {\tt hyperz} produced an excellent $\chi$$^{2}$ fit.
\cite{Mobasher2005} claimed the detection of a strong Balmer break between {\it H-band} and the {\it Spitzer} IRAC1
(3.6 $\micron$) bands. Simply selecting the best-fit reduced $\chi$$^{2}$\ model led the authors to conclude that
the object was a massive early-type galaxy  $z=6.5$, a stark contrast to the types of systems that should exist at this epoch. Their
results raised serious questions regarding the viability of the $\Lambda$-CDM paradigm
of building bigger galaxies from smaller building blocks.   The next best-fit model, from a secondary $\chi$$^{2}$ minimum, was a ``dusty'' starburst at $z\sim2.5$ with $E(B-V)=0.7$.
This model was rejected because it was considered too dusty at this redshift to be physically real, even though the authors estimated the 
probability of this fit to be 15\%. Revised SED-fitting, using newer 16 and 22  $\micron$\ photometry obtained with the IRS instrument on {\em Spitzer}, produced
a best-fit for a $z=1.7$ luminous infrared galaxy with a significant ($A_{\rm V}\sim$3 -- 4) amount of dust \citep{Chary2007}.
Similarly, \cite{Henry2008} claimed the discovery of a Lyman break galaxy
at $z\sim 9$ from NICMOS/{\it HST} imaging (based on a {\it F110W} dropout) and {\em Spitzer} imaging.  
Their $\chi$$^{2}$ minimization fitting rejected local minima models at $z\sim$2 -- 3
(produced by the degeneracy among the parameters).  Subsequently, it was observations from
MMT/Megacam blueward of the claimed break ({\it i}$^{'}$ $=$ 26 mag) that demonstrated 
the intermediate redshift solutions that were initially rejected were actually 
correct \citep{Henry2009}. Finally, \cite{Capak2011} presented three {\it z}-band
dropout candidates in which $\chi$$^{2}$ minimization fitting produced best solutions at
z $>$ 7, only to find that followup imaging of one blueward of the claimed break and at 
24 $\micron$ better supported at $z\sim1.7$ solution; and a second with observed 
{\it H}-band flux (from NIRSPEC/Keck spectroscopy) weaker than predicted by SED models.  \\
The authors of the works described above did not imply that low redshift solutions were entirely excluded, 
but selected high-$z$\ solutions based primarily on the $\chi$$^{2}$. Determining the accurate probability of these objects 
being at low redshift is therefore important. \\

\indent Any SED fitting effort is further 
complicated by the assumption that model galaxy spectra constructed from our knowledge 
of how stars evolve in the local universe is equally applicable to the earliest epochs.  
One should also be cognizant of the fact that the rest-frame wavelength range sampled by a fixed set 
of photometric filters will span a narrower range
with increasing redshift.  For the most recent claims of galaxies at $8<z <12$
\citep[e.g.][]{Yan2010,Bowens2011,Ellis2012}, WFC3/IR covers only $\Delta\lambda\sim$ 560 and 390 \AA, 
(for $z=8$\ and $z=12$, respectively, and excluding the 1216\AA\ break itself).  Including
the less sensitive and lower resolution IRAC1 and IRAC2 channels only widens this
to $\Delta\lambda\sim$ 3700 and 2600 \AA\ for $z=8$  and 12, respectively. Strong statistical priors 
exist for only a handful of lensed galaxies \citep[e.g.][]{Coe2013}. Recent work from \citet{Atek2011}
furthermore shows that there exists a population of low redshift galaxies with strong emission lines that can
mimic the color-selection criteria used to select $z\sim8$\ galaxies.
The relative lack of 
information about high redshift galaxy candidates, as well as the low 
S/N of the observations produce a situation where using sound statistical 
methods is particularly important when trying to determine the redshifts of these objects. \\
\indent The latest claims of $z>8$  galaxies based on $\chi$$^{2}$ minimization SED
fitting have been presented by \citet{Bowens2011}, \citet{Yan2010}, and \citet{Ellis2012} 
using deep WFC3/IR images of the Hubble Ultra Deep Field (HUDF). The results from \citet{Ellis2012} 
are particularly interesting since this project adds additional data 
(GO 12498; PI: Ellis, HUDF12) to the HUDF09 field (GO 11563; PI: Illingworth) in the F105W 
and F160W filters as well as new observations with the F140W filter. This filter spans a 
wavelength range inaccessible from the ground. 
All but one of the sources listed in \citet{Bowens2011} and none of the sources 
listed in \citet{Yan2010} were confirmed to be at $z>8.5$ using these new deeper data.

The new 
near-infrared measurements from the HUDF12 are the deepest observations available to date 
and for the foreseeable future. When combined with ancillary data from the HDUF09, these 
data allow one to construct spectral energy distributions of these sources spanning from 
observed {\it B}-band (ACS {\it F435W}) to IRAC1 and IRAC2 channels 
(3.4 and 4.6 $\micron$, respectively).  The 13 objects listed
in \citet{Ellis2012} show large red colors consistent with strong spectral breaks. 
While followup spectroscopic observations have only been attempted on one of the targets, 
UDFy-38135539 \citep{Lehnert2010,Bunker2013} and will be discussed later, the objects have
all been fit using what is now considered the ``standard method'' of  minimized $\chi^2$\ 
SED fitting, yielding redshifts of 8 $<$ z $<$ 12.  Given the limitations
of this method, and the numerous examples of what later turned out 
to be incorrect redshift identifications, we are motivated to apply more robust Bayesian 
techniques to better constrain the physical parameters of galaxies observed using broad 
band filters.  Central to this effort is the need to derive realistic estimates of the 
uncertainties for each derived parameter. 

Thus, using the sources and fluxes, including Advanced Camera for Surveys (ACS), WFC3 and IRAC, provided
in \citet{Ellis2012}, and reference therein, for 13 candidate high-$z$\ 1216\AA\ break galaxies we employ
 \PMCMC, a Markov Chain Monte Carlo (MCMC) approach 
\citep[for details see ][]{Pirzkal2012a,Pirzkal2012b}. For the purpose of this paper we define low redshift to be $z < 4$\ and high redshift to be $z >4$.  
Further, we treat equally all 13 objects listed in \citet{Ellis2012} as potential high redshift candidates to be tested with \PMCMC.
We are not implying that  \PMCMC\ is the only
viable method to perform this type of analysis.  MCMC is a well established approach to this
type of problem. The marginal S/N of current  high redshift candidates observations
warrant more robust techniques than the common $\chi^2$ minimization technique and the reliance on 
a single ``best fit'' model provided by {\tt hyper}-derived SED fitting software.

\section{APPLIED TECHNIQUES: MINIMIZED $\chi^2$\ SED FITTING VERSUS \PMCMC}\label{pmc}
\indent Nominally, the ``standard method'' of minimized $\chi^2$\ SED fitting works by 
matching photometric fluxes to a pre-selected grid of model templates until the residuals 
between model and data are minimized.  However, the choice of input models, the number of 
parameters (and whether any are constrained), parameter degeneracy and if the models 
sufficiently sample the entire range of the physical parameter space all affect the outcome of the fit. 
Any and all of these issues can produce ``best-fit'' solutions that are unphysical (i.e. a very 
old object in a young universe as in \citet{Labbe2010} and \citet{Richard2011}).  Once the number
of model parameters to be fit exceeds three it becomes computationally inefficient to fit 
parameter grid SEDs, more so for errors because each parameter must be refit following
a pure Monte Carlo approach (e.g. each parameter requires several thousand additional
iterations).  All of the most recent {\it HST} based high-$z$\ candidate objects have relied 
on some modified implementation of the {\tt hyperz} photometric code, or similar  $\chi^2$\ SED 
minimization techniques \citep[e.g.][]{McLure2010, McLure2011,Bowens2011, Ellis2012}.
While {\tt hyperz} has proven to be capable for accurately constraining low-$z$\ photometric 
redshifts for large numbers of sources in deep, often crowded fields later confirmed
with spectroscopic followup, it was designed to estimate the {\it most likely} photometric 
redshift, not produce credible intervals.  Its treatment of the probability 
density functions (PDFs) is inadequate.   Based on the {\tt hyperz} Users Manual and
J.-M. Miralles (2013, private communication) it minimizes over the extra dimensions ($\mathbf{T}$) instead of 
marginalizing over them, therefore the PDF  returned by {\tt hyperz} is given by:
\begin{equation}
p_{\tt hyperz}(z) = \min_{\mathbf{T}}p(z,\mathbf{T})
\end{equation}

This approach does not account for the {\it volume} of the parameter space probed 
resulting in unreliable confidence intervals, particularly in the case of additional maxima.
While such a distribution may be somewhat useful, the credible intervals produced
by {\tt hyperz} should not be trusted (J.-M. Miralles 2013, private communication).\\
\indent In contrast, MCMC, and our implementation of it, \PMCMC, is a  
method which randomly samples the entire parameter space but does not probe the multidimensional 
region uniformly.  The posterior probability distribution function can be 
constructed by simply creating a histogram of the values of a given parameter in the MCMC 
chain from values taken after it converges \citep[see Section 2 in ][for more details]{Pirzkal2012a} 
. Just as in \citet{Pirzkal2012a} we quote both the 95$\%$ and 68$\%$ 
credible regions here (see Section 2.1 of \citet{Pirzkal2012a} for a discussion of 
95$\%$ versus 68$\%$ credible regions, which we take to be the 95$\%$ and 68$\%$ highest posterior 
density, or HPD, regions, respectively).  While computationally more expensive 
(yet more efficient), our current implementation of the MCMC SED fitting code, \PMCMC, has 
three main features: \\
1. It does not depend on a pre-defined input model parameter grid and allows for a 
computationally efficient exploration of the input parameter space.\\
2. the effect of nebular emission lines and nebular emission continuum can be included.\\
3. It allows one to derive a statistically valid posterior PDF for each of the input 
parameters by integrating over the remaining parameters: 
\begin{equation}
p(z) = \int p(z,\mathbf{T})\,d\mathbf{T}
\end{equation}
To properly assess the possibility of lower-redshift solutions, it is imperative that the 
volume of parameter space be taken into account by integrating over the additional degrees 
of freedom.\\
\indent We have applied \PMCMC\ to the sample of high redshift candidates listed in 
\citet{Ellis2012} because this sample is based on the deepest {\it HST} observations to
date and provides candidate objects that could be the faintest examples of Lyman break
galaxies. Here, the analysis of redshifts and other parameters using the photometric fluxes
from \citet{Ellis2012} are based upon using the use of stellar population models from
\citet[][BC03]{bc03}, \citet[][CB07]{cb07}, and \citet[][MA05]{ma05} in conjunction with 
\PMCMC.  The parameters investigated include: stellar mass; a broad range of extinction 
values ($0<A_{\rm V}<10$); metallicity ($0.005<Z<0.05$); and stellar population ages 
(limited only by the age of the universe at a given redshift). We adopted flat priors for 
these parameters. We choose flat priors as these most closely match those used by
\citet{Ellis2012}, but we allowed for larger values of extinction.  Using a flat prior for the redshift is a conservative choice 
because it does not predispose the likelihood towards low or high redshifts.  
A redshift prior heavily biased towards higher redshift would further reduce the likelihood of these
sources to be low redshift interlopers. 
The application of these models has also been tested with 
and without the effects of nebular continuum and emission lines (as parameterized by the 
escape fraction parameter).  Details regarding the application of continuum nebular 
emission and emission lines to \PMCMC\ can be found in Section 2.2 of \citet{Pirzkal2012a}.  
In addition to using single stellar population models (SSP), the BC03 models were 
also tested using an exponentially decaying star formation history (SFH) model (parameterized by 
$\tau$). We use these stellar populations because we are confident that \PMCMC\ handles
them properly and we have extensively tested them. They also match the stellar population models
used by \citet{Ellis2012} and we would like to be able to compare the results of applying an MCMC
approach without increasing the complexity of the comparison with previous results. We note that
some \citep[e.g.,][]{Papovich2011} have advocated the use of a rising exponential star formation rate which would be useful to include
in future analyses of these objects. Thus, there are seven discrete population models used with \PMCMC :\\
(1) AaSSP BC03 model with nebular contribution; (2) a SSP BC03 model without nebular 
contribution; (3) a BC03 model with $e^{-t/\tau}$ star-formation history and without nebular 
contribution; (4) a SSP CB07 model with nebular contribution; (5) a SSP CB07 model without
nebular contribution; (6) a SSP M05 model with nebular contribution; and (7) a SSP M05
model without nebular contribution. 

\section{RESULTS}\label{results}
\subsection{Redshift Constraints}\label{redshift_results}
\indent A subset of the stellar population parameters, obtained using \PMCMC\ and BC03 stellar 
population model with nebular emission, is presented in the first five columns of Table  
\ref{tablebc03b}. For the most part, the redshift, extinction, stellar ages, stellar 
masses and metallically ranges are uniform among the seven input models we used with 
\PMCMC.  Any particular outliers are noted in Table 1.\\
\indent The results using the BC03 are representative of what is derived for nearly 
all of the models.  Table \ref{tablebc03b} lists the 68\% and the 95\% credible intervals 
derived using \PMCMC.  Table 1 clearly shows that the 95\% credible regions do {\it not} 
strongly constrain the redshift ranges of most of the sources.  It is also clear from 
Table 1 that the 95\% credible intervals are much larger than the 68\% credible 
intervals which means that redshift posterior PDFs are {\it not} Gaussian.  \\
\indent A comparison 
of the quality of the fits (parameterized as the log likelihood computed by MCMC) from the most representative high redshift ($z>4$) solution 
and that of the most representative low redshift ($z<4$) solution, shows that  high redshift models {\it appear} to fit the observations better than 
lower redshift models. 
As a first step, we can examine whether high redshift models always fit the data {\em significantly} better than 
alternative low redshift models by computing the significance {\it p} of a likelihood 
ratio test.  This is a comparison between the log likelihood of the
best fitting (usually $z>4$) and best fitting low-$z$\ ($z<4$) models.  If {\it p} $>$ 0.05 then
the hypothesis that the high-$z$\ model fits the observation better than the low-$z$\ model is rejected at the 
$2\sigma$\ level. \\
\indent While we used stellar population models both with and without the effect of nebular lines, we find that when using BC03 models without nebular emission, 7 of the 13 sources are reasonably fit by low 
redshift models.  However, if nebular emission is included then {\it 12 out of 13} objects 
are reasonably fit (i.e. {\it p}$>$0.05) by low redshift models.  The full range of {\it p} values obtained from using 
different stellar population models are shown in Table \ref{tablebc03b}. Clearly, the 
quality of the fits is not enough to confidently rule out these sources as low redshift 
interlopers.\\
\indent  The MCMC methodology also allows us to compute the {\it actual} probability that 
a source is at $z<4$, $P(z<4)$, by integrating the posterior PDF of each object. The range of $P(z<4)$ values
across all seven stellar population models described above is shown in the last column of 
Table \ref{tablebc03b}.  
Low redshift solutions {\it cannot} be strongly excluded based on the observed 
photometric break for many of the objects listed in Table \ref{tablebc03b}. When examining the
results using all seven stellar population models, 
9 out of 13 objects cannot be ruled out (i.e. $P(z<4)>0.05$) as low redshift interlopers in the best case 
scenario (i.e. selecting the stellar population models that most favor high redshift 
solutions), and none out of 13 in the worst case scenario (i.e. selecting the stellar 
population models that most favor low redshift solutions).\\
\indent Low redshift ($z<4$) solutions can, at best, be excluded at the $2\sigma$\ confidence level
($P(z<4) < 0.05$)  for only four sources (UDF12-3947-8076, UDF12-3954-6284, UDFy-3779600 and UDFy-38135539). 
For each of the seven stellar population, the average probability for $P(z<4)$, averaged across all 13 sources, is remarkably 
consistent and
 ranges from $19\%$\ to $26\%$\ with a global average of 21$\%$. We conclude that $\approx21\%$ 
of the sources in Table \ref{tablebc03b} are therefore likely to be low redshift 
interlopers.\\
\indent As in the case of earlier attempts to detect high-$z$\ systems \citep[e.g.][]{Henry2009}, 
it turns out that the limiting sensitivities of fluxes {\it blueward} of the presumptive 
break (i.e. ACS bands in our case) are not sufficient to provide strong constraints for 
the redshifts proposed by simple minimized $\chi$$^{2}$ and ''best fit" model SED fitting.   
Similarly, the IRAC observations are too shallow to provide enough constraints on the 
rest frame optical light from most of these objects.\\
\indent In the case of four sources for which low-$z$\ solutions can be {\it statistically}
rejected, three (UDF12-3947-8076, UDFy-3779600 and UDFy-38135539) are y-band dropouts and the brightest 
objects in the sample. Thus, relative to the ACS and IRAC detection limits, they have
enough flux in the WFC3 bands to produce a change in flux strong enough to resemble
a 1216\AA\ decrement.  The third object, UDF12-3954-6284, is very faint and only
detected in {\it one} band.  By definition it is a marginally acceptable candidate
because it is only detected in one band \citep[as noted by][]{Ellis2012}.  It is possible
that the observed photometric break could be caused by a strong emission line at low-$z$\
(see Section \ref{discussion}).\\
\indent The ACS and IRAC detection limits are too high to unambiguously identify a 1216\AA\ 
decrement for the remaining nine objects in the sample. We estimate that the ACS and
IRAC detection limits are $\approx5$\ and $\approx3$\ times too high (respectively) 
to distinguish between a 1216\AA\ decrement and a Balmer break in sources this faint.
This remains the main limiting factor in securely identifying sources at $z>8.5$ using 
WFC3 observations of the HUDF.  \\
\indent In Figure \ref{fig1} we show the SEDs of  objects UDF12-3954-6284
and UDFy-37796000. The first example is meant to illustrate how too high of rest frame optical limits do not allow to reject low redshift
solutions. The second example shows a clear detection of a strong photometric break  as well as constraining 
limits in both the rest frame UV and optical bands. 
In the case of UDF12-3954-6284, the IRAC detection limits are clearly too high to rule
 out that this object is a low redshift interloper, if we allow for nebular emission. 
 The log likelihood of the high-$z$\ and low-$z$\ solutions for this object 
are 720.9 and 722.4 with a likelihood ratio test confidence of 0.08 which indicates that 
the high redshift model does not fit the observation significantly (i.e. 2$\sigma$) better 
than the low redshift model. While models lacking nebular emission clearly favor a high redshift solution for
this object, the break can clearly be reproduced by an emission line.

A low redshift solution is much less likely in the case of  the significantly brighter 
object UDFy-37796000 with a redshift of $\approx1.6$ and $A_{\rm V}=2.8$. The log likelihoods of the 
high and low redshift solutions are 722.6 and 717.5, respectively, resulting in a very low
likelihood ratio test confidence values. This case demonstrates how strong rest-frame near UV and optical
limits can help rule out a Balmer break.

\subsection{Secondary Parameters}
\indent In addition to redshift constraints \PMCMC\ was used to determine other parameters 
including:  stellar population ages; extinction; metallicity; and stellar mass.  In the 
cases where models included nebular continuum, the escape fractions were determined; and 
in the case of BC03 with an exponentially decaying starburst, value of $\tau$ were derived. 
The 95\%\ and 68\%\ credible regions estimates for these parameters are given in Table 
\ref{tablebc03b} for each object.  In the cases of three parameters:  metallicity, escape 
fraction, and $\tau$ the posterior PDFs were essentially flat.  This confirms results from 
\citet{Pirzkal2012a} where it was demonstrated that these parameters can only be 
constrained using very high precision photometry (i.e. better than 1\% level when using broad-band
photometry),  which is 
not the case for the HUDF12, and that simply increasing the number of broad-band filter observations is not
sufficient.\\
\indent The interdependence of the input model parameters is illustrated in Figure 
\ref{fig2} which shows the two dimensional distribution of the posterior PDF's
for UDF12-3895-7114. The statistically more likely regions are shown using
proportionally lighter shades. This figure illustrates the statistical complexity
of fitting high redshift sources to stellar population models, the non Gaussian nature
of many of the input model parameter PDF's,  as well as the low redshift solutions
which fit the observations. As suggested above, deeper ACS observations (an increase in 
sensitivity by a factor of 5 for the {\it F850LP} ACS filter) would make it possible to 
more confidently exclude many of the low redshift solutions. 

\section{DISCUSSION}\label{discussion}
\begin{figure}
\includegraphics[width=5.5in]{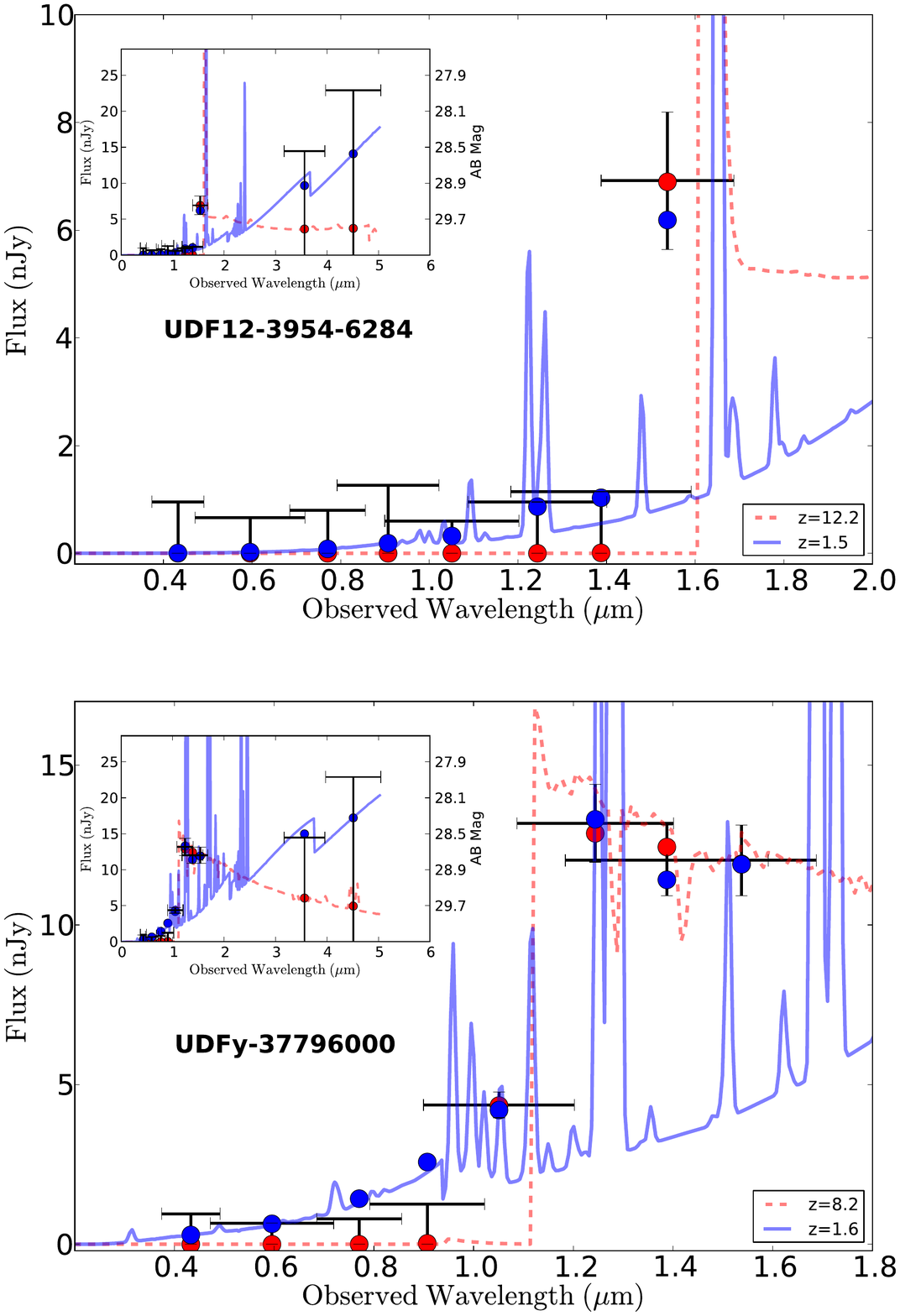}
\caption{High and low redshift solutions to the observed photometry of objects 
UDF12-3954-6284 and UDFy-37796000. The upper left sub-panels show all of the available 
photometry, including the infrared IRAC upper limits. The main panels shows the observed 
UV to NIR (ACS and WFC3 bands) region for the same object. Two models are shown in each 
case. The first, shown in a dashed red line, is a $z>8$ model while the solid blue line 
shows a much lower redshift solution at $z\approx2$. Observed fluxes are shown using error 
bars. Synthetic model fluxes in each observed filter are shown using large circles. 
\label{fig1}} 
\end{figure}

\begin{figure}
\includegraphics[width=7.0in]{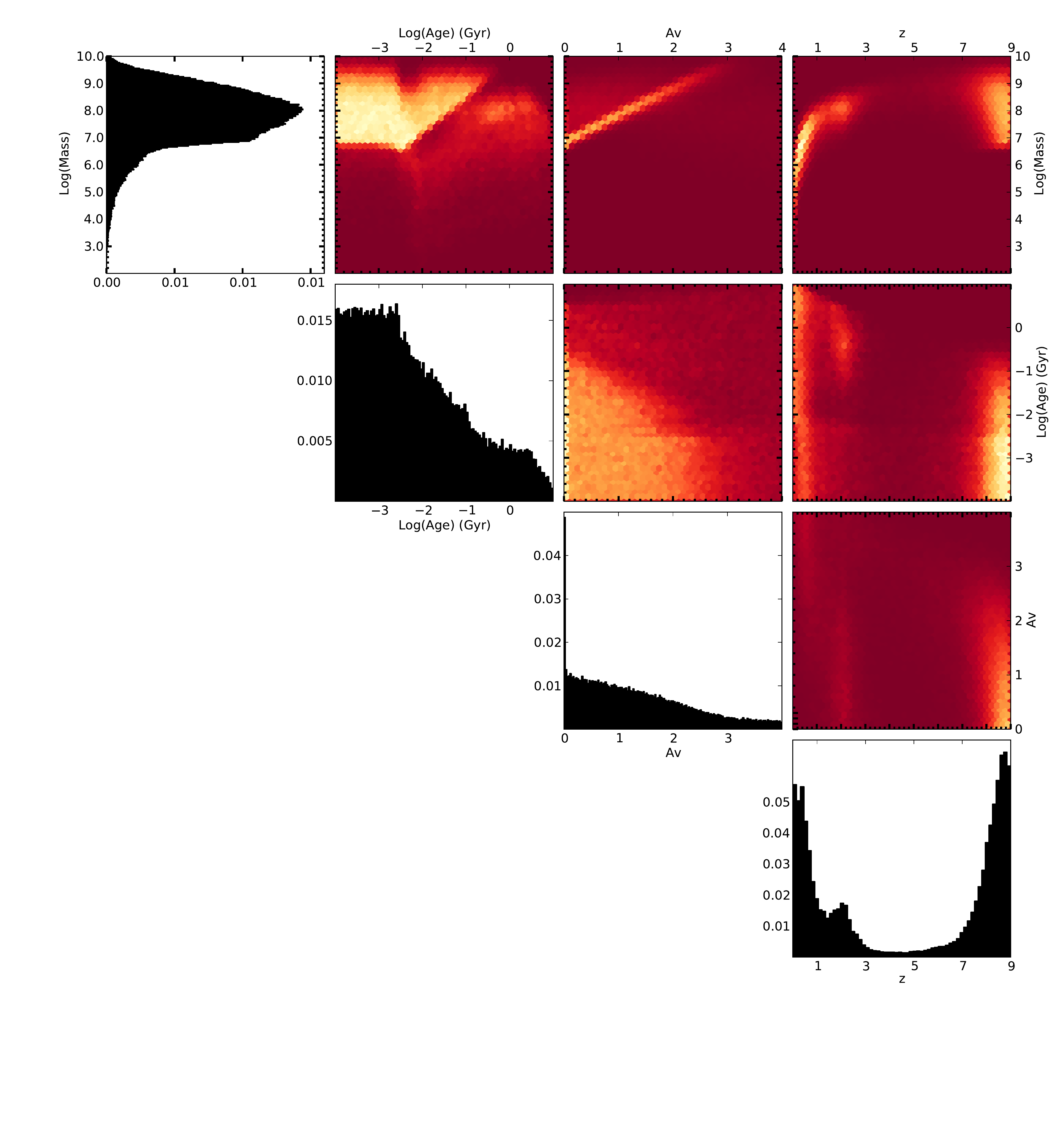}
\caption{Posterior probability density function estimates for object UDF12-3895-7114. The 
non-Gaussian, multi-modal nature of the redshift distribution is evident in the rightmost 
plots.  \label{fig2}}
\end{figure}

\indent  The low S/N and large errors associated with the ACS, WFC3, and {\em Spitzer}
observations of some of the faintest sources in the HUDF requires a more robust analysis than common $\chi^2$\ minimization 
techniques can provide. MCMC is able to provide this and allows us to 
derive accurate credible regions for redshifts, especially in the faint target, highest redshift regime. A complete SED analysis of the high redshift candidates 
discussed in this paper shows that there is a significant possibility ($\sim$1 in 5) that  
these sources are low to moderate redshift (z$<$4) interlopers. 
This is in contrast to the results presented by \citet{Ellis2012}, where they conclude that 7 out of 13 sources can 
be confidently classified as high-$z$\ galaxies.  Examining our 68\% redshift credible regions  might give the impression that our results agree well
with those of \citet{Ellis2012}. However, the limited number of 
measurements as well as the low S/N of the available measurements cause the redshift PDF of these sources to be highly non
Gaussian. This in turns lead us to revise upward the probability that these sources are low redshift interlopers. In the end, we find
 that only the redshifts of the brightest y-dropouts in this sample are likely to be at a high redshift ($z\approx8$).
Although the 128 orbits 
of the UDF12 have added considerable photometric depth to the UDF, the sources listed in 
Table \ref{fig1} are still extremely faint.  Moreover, given the large uncertainties in 
both flux and position, current IRAC observations do not strongly constrain the redshift 
of the vast majority of these objects.  Similarly, the ACS observations are of 
insufficient depth to strongly constrain the SED blue-ward of the (presumptive) Lyman 
(or possibly Balmer) break.  \\
\indent Spectroscopic observations, while difficult and time 
consuming, still remain a viable method for further constraining the proposed photometric 
redshifts and have been attempted for two of the sources in Table \ref{tablebc03b}.  The first, 
UDF12-3954-6284, was observed using the WFC3 G141 grism 
($\lambda$ $=$ 1.075-1.7 $\micron$ with $\delta$$\lambda$ $=$ 0.00465 $\micron$) as part 
of the 3D-HST survey (GO-12177; PI: van Dokkum) for two visits (4.7 ks each, the same 
orientation for each visit) and as a part of the CANDELS supernova follow-up program 
(GO-12099; PI: Riess) also for two visits
(6.6 ks and 15 ks) each with a different orientation.  \citet{Brammer2013} report
a 2.7$\sigma$ detection of an emission line at 1.599$\mu$m based on smoothing
and cross correlating the 2D spectrum with a kernel constructed from the central
0{\arcsec}.3 of the {\it F160W} thumbnail image of UDF12-3954-6284 and removing
any contamination using a contamination model (see their Figure 2).  The object is not 
detected in any single visit.  If the emission line were from Ly$\alpha$ it would
place the galaxy at $z\sim12$, but \citet{Brammer2013} reject this possibility based
not on the rest-frame equivalent width ($\sim$ 170 \AA) but because Ly$\alpha$ emission 
should be attenuated early in the epoch of re-ionization.  Instead, they conclude the
line is most likely [OIII], placing the object at $z\sim2.2$. Ironically, 
UDF12-3954-6284 was detected in only one WFC3 broad band filter and is one of the 
three objects with the strongest probabilities of being at high redshift.\\
\indent The second object with follow-up spectroscopic observations is UDFy-38135539,
all of which were obtained with ground-based instruments.
It was first observed with SINFONI on the VLT (integration time of 14.8 hr), 
centered on the {\it J}-band (1.1-1.4 $\micron$) resulting in a reported 6$\sigma$ 
detection of Ly$\alpha$, placing the galaxy at z $=$ 8.555 \citep{Lehnert2010}.  
Subsequent observations using X-SHOOTER (0.3-2.5 $\micron$) on the VLT and 
MOIRCS (0.9-1.78 $\micron$) on Subaru (5 hr and 11 hr integration times, 
respectively), have disputed this claim based on non-detections of Ly$\alpha$ \citep{Bunker2013}.  
They further noted that the {\it F105W} WFC3 photometry is inconsistent with both the line 
flux claimed by \citet{Lehnert2010} and the redshift from \citet{McLure2009}.  If the
redshift were correct, then the {\it F105W} flux should be detected at 15$\sigma$. 
The discrepancy between predicted F105W fluxes from SED-fitting (assuming high-$z$\ and a real line detection) 
and the actual measured F105W fluxes raises some concern regarding the true nature of this object.
\\

\indent Using \PMCMC\ we find that there are two populations of low-$z$\ galaxies which 
fit the available data as well as the high-$z$\ solutions.  The first are low-moderate 
dusty systems at $z\sim2$, the second are significantly dustier objects 
(up to {\it A}$_{\rm V}$ $=$ 4) at $z<1$.  Distinguishing between these low-$z$\ solutions
requires significantly more sensitive/deeper ACS and IRAC observations 
(see Figure \ref{fig1}) than are currently available (note: these would also serve to 
reject or improve the likelihood of high-$z$\ redshifts as well).  Deeper IRAC observations, should any be undertaken in the remaining time of the {\em Spitzer} Warm Mission, 
must contend with reaching source confusion limits as well as the higher
background noise from operating in ``warm mode.'' Contrary to common
wisdom and the conclusions reached by other high-$z$\ studies, deeper observations 
blue-ward (e.g {\it F850LP} or {\it F814W}) of the presumptive break {\it would} be extremely 
useful. (see Figure \ref{fig1}).  Alternatively, one can seek
sources that are brightened by lensing due to foreground galaxy clusters
to get more flux from high-$z$\ sources \citep[e.g.][]{Coe2013}. However, only a limited volume 
can be probed using this method.  \\
\indent Unfortunately, deeper observations in the optical bands using ACS 
are not likely considering the prohibitive observing times that would be required. 
Slitless spectroscopic observations using WFC3 with the G141 grism would provide an 
excellent method of rejecting or supporting the claims of high redshifts.  However, this 
requires multiple position angles to avoid contamination from other objects or 
observational artifacts in the field as well as observations significantly deeper than 
those attempted to date \citep{Pirzkal2012b}.  The G141 WFC3 grism observations would 
have the sensitivity to detect the emission lines potentially responsible for the observed 
photometric breaks. For example, in the case of object UDF12-3954-6284, an emission line 
with a flux as low as $\approx 1.5 \times 10^{-18}\ {\rm erg\ s^{-1}\ \AA^{-1}\ cm^{-2}}$\ could be 
solely responsible for a 29.3 mag detection in the F160W filter. Such a faint line 
would require upward of forty-two orbits to be {\em significantly} detected. 
In the absence of deeper HST or IRAC observations, observers will have to await the 
launch of {\em JWST}.  We estimate that 10ks observations at  $0.9\mu$m using NIRCAM with 
the {\em JWST} would  be deep enough to confirm that the observed photometric breaks are strong 
enough to be 1216\AA\ breaks. Until such observations are made, the nature as well as the 
volume density  of very high redshift candidates are likely to remain uncertain unless 
significantly deeper ACS or WFC3 observations are taken.  At the very least, our analysis 
suggests that the uncertainties of current estimates of the volume density of these 
objects should be adjusted to account for the possibility that nearly one in four of 
these objects is not a high redshift galaxy.

\clearpage
\pagestyle{empty}
\begin{deluxetable}{cccccccccc}
\tabletypesize{\footnotesize}
\setlength{\tabcolsep}{0.05in}
\tablehead{
\colhead{Object} &
\colhead{z} &
\colhead{log(Mass)} &
\colhead{$A_{\rm V}$} &
\colhead{$f_{\rm esc}$} &
\colhead{log(Age)} &
\colhead{$p$\tablenotemark{a}} &
\colhead{$P(z<4)$\tablenotemark{b}} \\
\colhead{} &
\colhead{} &
\colhead{$(M_{\sun})$} &
\colhead{} &
\colhead{$$} &
\colhead{(Gyr)} &
\colhead{(Min,Max)} &
\colhead{(Min,Max)}
}
\rotate
\startdata 
UDF12-3895-7114 & $ 8.5^{+2.3,+1.4}_{-2.2,-8.5} $ & $ 7.6^{+0.8,+1.9}_{-0.9,-1.8} $ & $ 0.8^{+0.7,+5.4}_{-0.8,-0.8} $ & $ 0.5^{+0.2,+0.5}_{-0.5,-0.5} $ & $ -2.3^{+0.7,+2.5}_{-1.7,-1.7} $ & 0.11,0.23 & 0.30,0.40  \\ 
UDF12-3921-6322 & $ 8.8^{+0.9,+0.9}_{-0.5,-8.7} $ & $ 7.5^{+0.7,+1.8}_{-0.6,-1.2} $ & $ 0.5^{+0.5,+5.3}_{-0.5,-0.5} $ & $ 0.5^{+0.2,+0.5}_{-0.5,-0.5} $ & $ -2.6^{+0.7,+2.6}_{-1.4,-1.4} $ & 0.01,0.05 & 0.06,0.21  \\ 
UDF12-3947-8076\tablenotemark{c}  & $ 8.6^{+0.8,+1.6}_{-0.4,-7.9} $ & $ 8.1^{+0.6,+1.1}_{-0.6,-1.1} $ & $ 0.6^{+0.3,+3.4}_{-0.6,-0.6} $ & $ 0.6^{+0.4,+0.4}_{-0.2,-0.5} $ & $ -2.3^{+0.5,+2.3}_{-1.7,-1.7} $ & 0.01,0.31 & 0.03,0.84  \\ 
UDF12-3954-6284\tablenotemark{d} & $ 12.0^{+0.5,+0.6}_{-0.3,-1.4} $ & $ 7.8^{+0.4,+1.4}_{-0.5,-0.7} $ & $ 0.0^{+0.2,+1.0}_{-0.0,-0.0} $ & $ 0.5^{+0.2,+0.5}_{-0.5,-0.5} $ & $ -2.9^{+0.5,+1.6}_{-1.1,-1.1} $ & 0.00,0.24 & 0.00,0.13  \\ 
UDF12-4106-7304 & $ 9.4^{+2.1,+1.7}_{-1.2,-9.4} $ & $ 7.7^{+0.8,+1.8}_{-0.9,-1.8} $ & $ 0.7^{+0.7,+5.8}_{-0.7,-0.7} $ & $ 0.5^{+0.5,+0.5}_{-0.2,-0.5} $ & $ -2.4^{+0.7,+2.6}_{-1.6,-1.6} $ & 0.06,0.35 & 0.12,0.28  \\ 
UDF12-4265-7049 & $ 9.1^{+1.3,+1.5}_{-0.9,-9.1} $ & $ 7.8^{+0.7,+1.9}_{-0.9,-1.5} $ & $ 0.6^{+0.6,+5.2}_{-0.6,-0.6} $ & $ 0.5^{+0.2,+0.5}_{-0.5,-0.5} $ & $ -2.4^{+0.6,+2.4}_{-1.6,-1.6} $ & 0.06,0.19 & 0.18,0.26  \\ 
UDF12-4344-6547 & $ 8.7^{+0.9,+1.1}_{-0.6,-8.5} $ & $ 7.5^{+0.7,+1.9}_{-0.8,-1.2} $ & $ 0.5^{+0.5,+4.8}_{-0.5,-0.5} $ & $ 0.5^{+0.2,+0.5}_{-0.5,-0.5} $ & $ -2.6^{+0.6,+2.3}_{-1.4,-1.4} $ & 0.03,0.13 & 0.14,0.23  \\ 
UDFj-35427336 & $ 3.4^{+5.5,+6.6}_{-3.4,-3.4} $ & $ 7.7^{+1.1,+1.8}_{-0.8,-2.6} $ & $ 1.7^{+1.9,+6.0}_{-1.7,-1.7} $ & $ 0.5^{+0.2,+0.5}_{-0.5,-0.5} $ & $ -2.0^{+0.8,+2.5}_{-2.0,-2.0} $ & 0.16,0.47 & 0.40,0.50  \\ 
UDFj-38116243 & $ 9.6^{+2.4,+1.9}_{-1.3,-9.6} $ & $ 7.8^{+0.8,+2.0}_{-1.0,-1.8} $ & $ 0.9^{+0.7,+5.7}_{-0.9,-0.9} $ & $ 0.5^{+0.2,+0.5}_{-0.5,-0.5} $ & $ -2.3^{+0.7,+2.4}_{-1.7,-1.7} $ & 0.13,0.50 & 0.20,0.30  \\ 
UDFj-43696407 & $ 2.1^{+6.7,+9.0}_{-2.1,-2.1} $ & $ 7.5^{+1.2,+1.9}_{-0.7,-2.5} $ & $ 2.0^{+1.8,+5.4}_{-2.0,-2.0} $ & $ 0.5^{+0.2,+0.5}_{-0.5,-0.5} $ & $ -2.0^{+0.9,+2.5}_{-2.0,-2.0} $ & 0.78,1.00 & 0.41,0.92  \\ 
UDFy-33436598 & $ 7.8^{+0.7,+0.9}_{-0.4,-7.3} $ & $ 7.7^{+0.4,+1.3}_{-0.7,-0.9} $ & $ 0.6^{+0.3,+2.9}_{-0.6,-0.6} $ & $ 0.5^{+0.5,+0.5}_{-0.2,-0.5} $ & $ -2.6^{+0.6,+2.5}_{-1.4,-1.4} $ & 0.01,0.05 & 0.09,0.32  \\ 
UDFy-37796000\tablenotemark{e} & $ 8.1^{+0.2,+0.5}_{-0.2,-6.6} $ & $ 7.7^{+0.3,+0.8}_{-0.4,-0.6} $ & $ 0.3^{+0.2,+2.7}_{-0.3,-0.3} $ & $ 0.6^{+0.4,+0.4}_{-0.2,-0.5} $ & $ -2.9^{+0.4,+1.3}_{-1.1,-1.1} $ & 0.00,0.00 & 0.00,0.08  \\ 
UDFy-38135539\tablenotemark{c}  & $ 8.4^{+0.2,+0.3}_{-0.2,-0.3} $ & $ 8.1^{+0.4,+0.8}_{-0.5,-0.7} $ & $ 0.5^{+0.2,+0.8}_{-0.5,-0.5} $ & $ 0.7^{+0.3,+0.3}_{-0.2,-0.6} $ & $ -2.6^{+0.4,+1.1}_{-1.4,-1.4} $ & 0.00,0.69 & 0.00,0.13  \\ 
Average & & & & & & & 0.21 \\ 

\enddata
\caption{Derived physical properties of $z>8$ candidates in the HUDF}
\tablecomments{This table lists the median values for each parameter as well as the 68\% and 95\% credible regions (highest posterior density regions), separated by a comma,\label{tablebc03b} when using BC03 stellar population models.\\
a - Likelihood ratio test significance when comparing the quality of the best fitting model to the quality of the best fit at $z<4$. Significance values greater than 0.05 ($2\sigma$) imply that high redshift models do not fit the data significantly better. The minimum and maximum values of $p$ across all seven different input models are shown.\\
b - Probability that this object is at $z<4$. The minimum and maximum values of $P(z<4)$ across all seven different input models are shown.\\
c - MA05 models without nebular emission reject all low z solutions for this object.\\
d - BC03 models with exponentially decaying SFH and CB07 models do not reject low redshift solutions. The nature of this object is further discussed in Section \ref{results}.\\
e - CB07 and MA05 models lead to the rejection of low redshift solutions.\\
}
\end{deluxetable}

\clearpage

\section{Acknowledgment}
We would like to warmly thank J. M. Miralles for a fruitful discussion of the working of  
{\tt hyperz}. We would also like to thank F. Pierfederici for his contributions that 
ultimately allowed us to run \PMCMC\ on distributed computers using Celery and HT-Condor.
Finally, we  thank Dr. Richard Ellis, Dr. Ross McLure and Dr. Rychard Bouwens and their collaborators for 
constructive and insightful comments.


\begin{thebibliography}{}
\bibitem[Atek et al.(2011)]{Atek2011} Atek, H., Siana, B., Scarlata, C., et al.\ 2011, \apj, 743, 121 
\bibitem[Bolzonella et al.(2000)]{Bolzonella2000} Bolzonella, M., Miralles, J.-M., \& Pell{\'o}, R.\ 2000, \aap, 363, 476 
\bibitem[Bouwens et al.(2011)]{Bowens2011} Bouwens, R.~J., Illingworth, G.~D., Labbe, I., et al.\ 2011, \nat, 469, 504 
\bibitem[Brammer et al.(2013)]{Brammer2013} Brammer, G.~B., van Dokkum, P.~G., Illingworth, G.~D., et al.\ 2013, ApJL, 765, L2 
\bibitem[Bremer et al.(2004)]{Bremer2004} Bremer, M.~N., Jensen, J.~B., Lehnert, M.~D., et al. \ 2004, ApJL, 615, L1 
\bibitem[Bruzual \& Charlot(2003)]{bc03} Bruzual, G., \& Charlot, S.\ 2003, \mnras, 344, 1000 
\bibitem[Bunker et al.(2013)]{Bunker2013} Bunker, A.~J., Caruana, J., Wilkins, S.~M., et al.\ 2013, \mnras, 430, 3314
\bibitem[Capak et al.(2011)]{Capak2011} Capak, P., Mobasher, B., Scoville, N.~Z., et al. \ 2011, \apj, 730, 68
\bibitem[Charlot and Bruzual(2013)]{cb07} Charlot, S., Bruzual, G., 2013, in prep
\bibitem[Chary et al.(2007)]{Chary2007} Chary, R.~R., Teplitz, H.~I., Dickinson, M.~E. et al. \ 2007, \apj, 665, 257
\bibitem[Coe et al.(2013)]{Coe2013} Coe, D., Zitrin, A., Carrasco, M., et al.\ 2013, \apj, 762, 32 
\bibitem[Dickinson et al.(2004)]{Dickinson2004} Dickinson, M., Stern, D., Giavalisco, H.~C. et al. \ 2004, ApJL, 600, 99
\bibitem[Ellis et al.(2013)]{Ellis2012} Ellis, R.~S., McLure, R.~J., Dunlop, J.~S., et al.\ 2013, ApJL, 763, L7 
\bibitem[Fan et al.(2006)]{Fan2006} Fan, X., Strauss, M.~A., Becker, R.~H., et al.\ 2006, \aj, 132, 117 
\bibitem[Henry et al.(2008)]{Henry2008} Henry, A.~L., Malkan, M.~A., Colbert, J.~W., et al. \ 2008, ApJL, 680, L97
\bibitem[Henry et al.(2009)]{Henry2009} Henry, A.~L., Siana, B., Malkan, M.~A. et al. \ 2009, \apj, 697, 1128
\bibitem[Hu et al. (2004)]{Hu2004} Hu, E.~M., Cowie, L.~L., Capak, P. et al. \ 2004, \aj, 127, 563
\bibitem[Komatsu et al.(2011)]{Komatsu2011} Komatsu, E., Smith, K.~M., Dunkley, J., et al.\ 2011, \apjs, 192, 18 
\bibitem[Labb\'{e} et al.(2003)]{Labbe2003} Labb\'{e}, I, Franx, M., Rudnick, G. et al. \ 2003, \aj, 125, 1107
\bibitem[Labb\'{e} et al.(2010)]{Labbe2010} Labb\'{e}, I, Gonzalez, V., Bouwens, R.~J. et al. \ 2010, \apj, 716, 103
\bibitem[Lehnert et al.(2005)]{Lehnert2005} Lehnert, M.~D., F\"{o}rster-Schreiber, N.~M., \& Bremer, M.~N. \ 2005, \apj, 624, 80
\bibitem[Lehnert et al.(2010)]{Lehnert2010} Lehnert, M.~D., Nesvadba, N.~P.~H., Cuby, J.~G., et al. \ 2010, \nat, 467, 940
\bibitem[Malhotra et al.(2001)]{Malhotra2001} Malhotra, S., Rhoads, J., Dey, A., Stern, D., \& Spinrad, H.\ 2001, 
in ASP Conf. Proc. 240, Gas and Galaxy Evolution, ed. J. E. Hibbard, M. Rupen, \& J. H. van Gorkom (San Francisco, CA: ASP), 97
\bibitem[Malhotra et al.(2005)]{Malhotra2005} Malhotra, S., Rhoads, J.~E., Pirzkal, N., et al.\ 2005, \apj, 626, 666 
\bibitem[Maraston(2005)]{ma05} Maraston, C.\ 2005, \mnras, 362, 799 
\bibitem[McLure et al.(2009)]{McLure2009} McLure, R.~J., Cirasuolo, M., Dunlop, J.~S. et al. \ 2009, \mnras, 395, 2196
\bibitem[McLure et al.(2010)]{McLure2010} McLure, R.~J., Dunlop, J.~S., Cirasuolo, M., et al.\ 2010, \mnras, 403, 960 
\bibitem[McLure et al.(2011)]{McLure2011} McLure, R.~J., Dunlop, J.~S., de Ravel, L., et al.\ 2011, \mnras, 418, 2074 
\bibitem[Mobasher et al.(2005)]{Mobasher2005} Mobasher, B., Dickinson, M., Ferguson, H.~C. et al. \ 2005, \apj, 635, 832
\bibitem[Papovich et al.(2001)]{Papovich2001} Papovich, C., Dickinson, M., Ferguson, H.~C. et al. \ 2001, \apj, 559, 620
\bibitem[Papovich et al.(2011)]{Papovich2011} Papovich, C., Finkelstein, S.~L., Ferguson, H.~C., Lotz, J.~M., \& Giavalisco, M.\ 2011, \mnras, 412, 1123 
\bibitem[Pell\'{o} et al. (2004)]{Pello2004} Pell\'{o}, R., Schaerer, D., Richard, J., et al. \ 2004, A\&A, 416, L35
\bibitem[Pirzkal et al.(2007)]{Legos2007} Pirzkal, N., Malhotra, S., Rhoads, J.~E., \& Xu, C. \ 2007, \apj, 667, 49
\bibitem[Pirzkal et al.(2012a)]{Pirzkal2012a} Pirzkal, N., Rothberg, B., Nilsson, K.~K., et al.\ 2012, \apj, 748, 122 
\bibitem[Pirzkal et al.(2012b)]{Pirzkal2012b} Pirzkal, N., Rothberg, B., Chun, L., et al.\ 2012, \apj, accepted (arxiv:1208.5535)
\bibitem[Pirzkal et al.(2004)]{Pirzkal2004} Pirzkal, N., Xu, C., Malhotra, S., et al.\ 2004, \apjs, 154, 501 
\bibitem[Richard et al.(2011)]{Richard2011} Richard, J., Kneib, J.~P., Ebeling, H. et al. \ 2011, \mnras, 414, L31
\bibitem[Rhoads et al.(2009)]{Rhoads2009} Rhoads, J.~E., Malhotra, S., Pirzkal, N., et al.\ 2009, \apj, 697, 942
\bibitem[Schaerer \& Pell\'{o} (2005)]{Schaerer2005} Schaerer, D., \& Pell\'{o}, R. \ 2005, \mnras, 362, 1054
\bibitem[Songaila \& Cowie(2002)]{Songaila2002} Songaila, A., \& Cowie, L.~L.\ 2002, \aj, 123, 2183 
\bibitem[Stanway et al.(2008)]{Stanway2008} Stanway, E.~R., Bremer, M.~N., \& Lehnert, M.~D.\ 2008, \mnras, 385, 493 
\bibitem[Thompson et al.(2001)]{Thomspon2001} Thompson, R.~I., Weymann, R.~J., \& Storrie-Lombardi, L.~J. \ 2001, \apj, 546, 694
\bibitem[Weatherley, et al.(2004)]{Weatherley2004} Weatherley, S.~J., Warren, S.~J., \& Babbedge, T.~S.~R. \ 2004, A\&A, 428, L29
\bibitem[Yan et al.(2010)]{Yan2010} Yan, H.-J., Windhorst, R.~A., Hathi, N.~P., et al.\ 2010, RAA, 10, 867 
\end{thebibliography}
\end{document}